\documentclass[a4paper, floatfix, nofootinbib, twocolumn, longbibliography, showpacs]{revtex4-1}

\usepackage[utf8]{inputenc}
\usepackage{epsfig}
\usepackage[T1]{fontenc}
\usepackage[english]{babel}
\usepackage[table]{xcolor}
\usepackage{t1enc}
\usepackage{graphicx}
\usepackage{amssymb}
\usepackage{amsmath}
\usepackage{relsize}
\usepackage{overpic}
\usepackage{bm}
\usepackage{hyperref}

\usepackage[normalem]{ulem}

\newcommand{\avg}[1]{{\left<#1\right>}}

\usepackage{mathtools}
\def\multiset#1#2{\ensuremath{\left(\kern-.3em\left(\genfrac{}{}{0pt}{}{#1}{#2}\right)\kern-.3em\right)}}

\usepackage{amsmath}

\usepackage{todonotes}
\usepackage{lipsum}

\usepackage{verbatim}
\usepackage{overpic}
\usepackage{booktabs}
\usepackage{placeins}

\begin{document}

\title{Network structure, metadata and the prediction of missing nodes
and annotations}

\author{Darko Hric}
\email{darko.hric@aalto.fi}
\affiliation{Department of Computer Science, Aalto University School of
Science, P.O.  Box 12200, FI-00076 Aalto, Finland}

\author{Tiago P. Peixoto}
\email{t.peixoto@bath.ac.uk}
\affiliation{Department of Mathematical Sciences and Centre for Networks
and Collective Behaviour, University of Bath, Claverton Down, Bath BA2
7AY, United Kingdom}
\affiliation{ISI Foundation, Via Alassio 11/c, 10126, Turin, Italy}
\affiliation{Institut f\"ur Theoretische Physik, Universit\"at Bremen,
  Hochschulring 18, D-28359 Bremen, Germany}

\author{Santo Fortunato}
\email{santo.fortunato@aalto.fi}
\affiliation{Department of Computer Science, Aalto University School of
Science, P.O.  Box 12200, FI-00076, Finland}
\affiliation{Center for Complex Networks and Systems Research, School of
Informatics and Computing, Indiana University, Bloomington, USA}


\begin{abstract}
  The empirical validation of community detection methods is often based
  on available annotations on the nodes that serve as putative
  indicators of the large-scale network structure. Most often, the
  suitability of the annotations as topological descriptors itself is
  not assessed, and without this it is not possible to ultimately
  distinguish between actual shortcomings of the community detection
  algorithms on one hand, and the incompleteness, inaccuracy or
  structured nature of the data annotations themselves on the other. In
  this work we present a principled method to access both aspects
  simultaneously. We construct a joint generative model for the data and
  metadata, and a nonparametric Bayesian framework to infer its
  parameters from annotated datasets. We assess the quality of the
  metadata not according to its direct alignment with the network
  communities, but rather in its capacity to predict the placement of
  edges in the network. We also show how this feature can be used to
  predict the connections to missing nodes when only the metadata is
  available, as well as missing metadata. By investigating a wide range
  of datasets, we show that while there are seldom exact agreements
  between metadata tokens and the inferred data groups, the metadata is
  often informative of the network structure nevertheless, and can
  improve the prediction of missing nodes. This shows that the method
  uncovers meaningful patterns in both the data and metadata, without
  requiring or expecting a perfect agreement between the two.
\end{abstract}

\maketitle

\section{Introduction}

The network structure of complex systems determine their function and
serve as evidence for the evolutionary mechanisms that lie behind
them. However, very often their large-scale properties are not directly
accessible from the network data, and need to be indirectly derived via
nontrivial methods. The most prominent example of this is the task of
identifying modules or ``communities'' in networks, that has driven a
substantial volume of research in recent
years~\cite{fortunato_community_2010,porter_communities_2009,newman_communities_2011}. Despite
these efforts, it is still an open problem both how to characterize such
large-scale structures and how to effectively detect them in real
systems. In order to assist in bridging this gap, many researchers have
compared the features extracted from such methods with known information
--- metadata, or ``ground truth'' --- that putatively correspond to the
main indicators of large-scale
structure~\cite{yang_defining_2012,yang_community-affiliation_2012,yang_structure_2012}.
However, this assumption is often accepted at face value, even when such
metadata may contain a considerable amount of noise, is incomplete or is
simply irrelevant to the network structure. Because of this, it is not
yet understood if the discrepancy observed between the metadata and the
results obtained with community detection
methods~\cite{yang_defining_2012,hric_community_2014} is mainly due to
the ineffectiveness of such methods, or to the lack of correlation
between the metadata and actual structure.

In this work, we present a principled approach to address this issue.
The central stance we take is to make no fundamental distinction between
data and metadata, and construct generative processes that account for
both simultaneously. By inferring this joint model from the data and
metadata, we are able to precisely quantify the extent to which the data
annotations are related to the network structure, and vice
versa\footnote{Here we consider exclusively annotation on the
nodes. Networks may also possess annotations on the edges, which may be
treated as edge covariates or layers, as already considered extensively
in the literature
(e.g.~\cite{peixoto_inferring_2015,stanley_clustering_2015,aicher_learning_2015,valles-catala_multilayer_2016}).}. This
is different from approaches that explicitly assume that the metadata
(or a portion thereof) are either exactly or approximately correlated
with the best network
division~\cite{moore_active_2011,leng_active_2013,peel_active_2015,
yang_community_2013,bothorel_clustering_2015,zhang_phase_2014,zhang_community_2013,freno_learning_2011}.
With our method, if the metadata happens to be informative on the
network structure, we are able to determine how; but if no correlation
exists between the two, this gets uncovered as well. Our approach is
more in line with a recent method by Newman and
Clauset~\cite{newman_structure_2016} --- who proposed using available
metadata to guide prior probabilities on the network partition
--- but here we introduce a framework that is more general in three
important ways: Firstly, we do not assume that the metadata is present
in such a way that it corresponds simply to a partition of the
nodes. While the latter can be directly compared to the outcome of
conventional community detection methods, or used as priors in the
inference of typical generative models, the majority of datasets contain
much richer metadata, where nodes are annotated multiple times, with
heterogeneous annotation frequencies, such that often few nodes possess
the exact same annotations. Secondly, we develop a nonparametric
Bayesian inference method that requires no prior information or \emph{ad
hoc} parameters to be specified, such as the number of communities. And
thirdly, we are able not only to obtain the correlations between
structure and annotations based on statistical evidence, but also we are
capable of assessing the metadata in its \emph{power to predict the
network structure}, instead of simply their correlation with latent
partitions. This is done by leveraging the information available in the
metadata to predict \emph{missing nodes} in the network. This contrasts
with the more common approach of predicting missing
edges~\cite{liben-nowell_link-prediction_2007,clauset_hierarchical_2008,guimera_missing_2009,
  guimera_network_2013,rovira-asenjo_predicting_2013,musmeci_bootstrapping_2013,
  cimini_estimating_2015},
which cannot be used when entire nodes have not been observed and need
to be predicted, and with other approaches to detect missing nodes,
which are either heuristic in nature~\cite{rossi_transforming_2012}, or
rely on very specific assumptions on the data generating
process~\cite{bringmann_learning_2010,kim_network_2011}.  Furthermore,
our method is also capable of clustering the metadata themselves,
separating them in equivalence classes according to their occurrence in
the network. This clustering of the metadata is done simultaneously with
the clustering of the network data itself, with both aspects aiding each
other, and thus providing a full generalization of the task of community
detection for annotated networks. As we show, both features allows us to
distinguish informative metadata from less informative ones, with
respect to the network structure, as well as to predict missing
annotations.

In the following we describe our method and illustrate its use with some
examples based on real data. We then follow with a systematic analysis
of many empirical datasets, focusing on the prediction of nodes from
metadata alone. We show that the predictiveness of network structure
from metadata is widely distributed --- both across and within datasets
--- indicating that typical network annotations vary greatly in their
connection to network structure.

\section{Joint model for data and metadata}

\begin{figure}
  \includegraphics{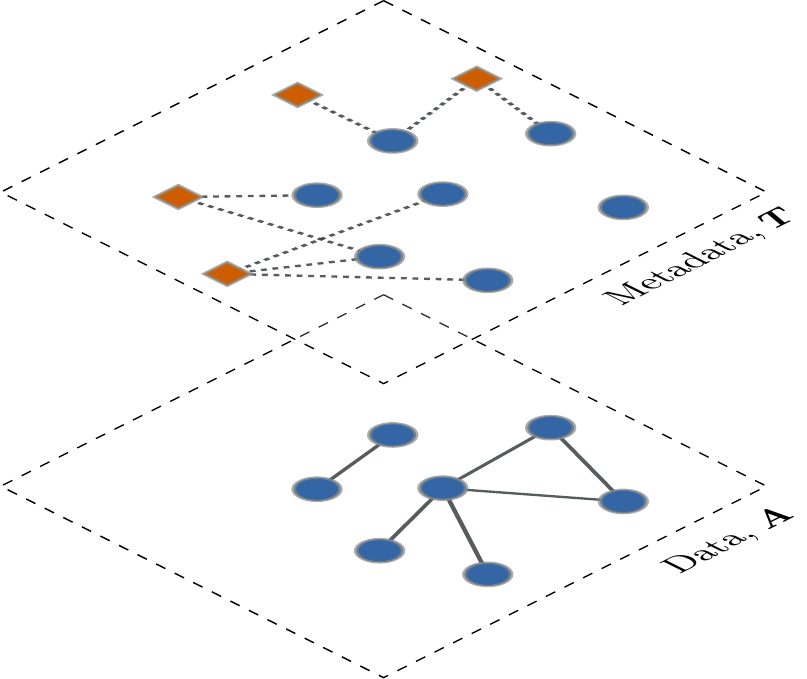}
    \caption{Schematic representation
    of the joint data-metadata model. The data layer is composed of data
    nodes and is described by an adjacency matrix $\bm{A}$, and the
    metadata layer is composed of the same data nodes, as well as tag
    nodes, and is described by a bipartite adjacency matrix
    $\bm{T}$. Both layers are generated by two coupled degree-corrected
    SBMs, where the partition of the data nodes into groups is the same
    in both layers. \label{fig:diagram}}
\end{figure}

Our approach is based on a unified representation of the network data
and metadata. We assume here the general case where the metadata is
discrete, and may be arbitrarily associated with the nodes of the
network. We do so by describing the data and metadata as a single graph
with two node and edge types (or
\emph{layers}~\cite{kivela_multilayer_2014,de_domenico_mathematical_2013}),
as shown in Fig.~\ref{fig:diagram}. The first layer corresponds to the
network itself (the ``data''), where an edge connects two ``data''
nodes, with an adjacency matrix $\bm{A}$, where $A_{ij}=1$ if an edge
exists between two data nodes $i$ and $j$, or $A_{ij}=0$ otherwise. This
layer would correspond to the entire data if the metadata were to be
ignored. In the second layer both the data and the metadata nodes are
present, and the connection between them is represented by a bipartite
adjacency matrix $\bm{T}$, where $T_{ij}=1$ if node $i$ is annotated
with a metadata token $j$ (henceforth called a \emph{tag} node), or
$T_{ij}=0$ otherwise. Therefore, a single data node can be associated
with zero, one or multiple tags, and likewise a single tag node may be
associated with zero, one or multiple data nodes. Within this general
representation we can account for a wide spectrum of discrete node
annotations. In particular, as it will become clearer below, we make no
assumption that individual metadata tags actually correspond to specific
disjoint groups of nodes.

We construct a generative model for the matrices $\bm{A}$ and $\bm{T}$
by generalizing the hierarchical stochastic block model
(SBM)~\cite{peixoto_hierarchical_2014} with
degree-correction~\cite{karrer_stochastic_2011} for the case with edge
layers~\cite{peixoto_inferring_2015}. In this model, the nodes and tags
are divided into $B_d$ and $B_t$ groups, respectively. The number of
edges between data groups $r$ and $s$ are given by the parameters
$e_{rs}$ (or twice that for $r=s$), and between data group $r$ and tag
group $u$ by $m_{ru}$. Both data and tag nodes possess fixed degree
sequences, $\{k_i\}$ and $\{d_i\}$, for the data and metadata layers,
respectively, corresponding to an additional set of parameters. Given
these constraints, a graph is generated by placing the edges randomly in
both layers independently, with a joint likelihood
\begin{equation}\label{eq:likelihood}
  P(\bm{A}, \bm{T}| \bm{b}, \theta, \bm{c}, \gamma) = P(\bm{A} | \bm{b}, \theta) P(\bm{T} | \bm{b},  \bm{c}, \gamma),
\end{equation}
where $\bm{b}=\{b_i\}$ and $\bm{c}=\{c_i\}$ are the group memberships of
the data and tag nodes, respectively, and both $\theta=(\{e_{rs}\},
\{k_i\})$ and $\gamma=(\{m_{ru}\}, \{d_i\})$ are shorthands for the
remaining model parameters in both layers. Inside each layer, the
log-likelihood is\footnote{Eq.~\ref{eq:dc-likelihood} is an approximation
that is valid for sparse graphs, where the occurrence of parallel edges
can be neglected. If this is not the case, the likelihood should be
appropriately modified. See
Refs.~\cite{peixoto_entropy_2012,peixoto_model_2015} for more
details.}~\cite{karrer_stochastic_2011,peixoto_entropy_2012}
\begin{equation}\label{eq:dc-likelihood}
  \ln P(\bm{A} | \bm{b}, \theta) \approx -E - \frac{1}{2}\sum_{rs}e_{rs}\ln\frac{e_{rs}}{e_re_s}-\sum_i\ln k_i!,
\end{equation}
and analogously for $P(\bm{T} | \bm{b}, \bm{c}, \gamma)$. Since the data
nodes have the same group memberships in both layers, this provides a
coupling between them, and we have thus a joint model for data and
metadata. This model is general, since it is able to account
simultaneously for the situation where there is a perfect correspondence
between data and metadata (for example, when $B_d=B_t$ and the matrix
$m_{ru}$ connects one data group to only one metadata group), when the
correspondence is non-existent (the matrix $\bm{T}$ is completely
random, with $B_t=1$), as well as any elaborate relationship between
data and metadata in between. In principle, we could fit the above model
by finding the model parameters that maximize the likelihood in
Eq.~\ref{eq:likelihood}. Doing so would uncover the precise relationship
between data and metadata under the very general assumptions taken
here. However, for this approach to work, we need to know \emph{a
priori} the number of groups $B_d$ and $B_t$. This is because the
likelihood of Eq.~\ref{eq:likelihood} is parametric (i.e. it depends on
the particular choices of $\bm{b}$, $\bm{c}$, $\theta$ and $\gamma$),
and the degrees of freedom in the model will increase with $B_d$ and
$B_t$. As the degrees of freedom increase, so will the likelihood, and
the perceived quality of fit of the model. If we follow this criterion
blindly, we will put each node and metadata tag in their individual
groups, and our matrices $e_{rs}$ and $m_{rs}$ will correspond exactly
to the adjacency matrices $\bm{A}$ and $\bm{T}$, respectively. This is
an extreme case of \emph{overfitting}, where we are not able to
differentiate random fluctuations in data from actual structure that
should be described by the model. The proper way to proceed in this
situation is to make the model \emph{nonparametric}, by including
noninformative Bayesian priors on the model parameters $P(\bm{b})$,
$P(\bm{c})$, $P(\theta)$ and $P(\gamma)$, as described in
Ref.~\cite{peixoto_hierarchical_2014, peixoto_model_2015} (See also
Appendix~\ref{app:priors}). By maximizing the joint nonparametric
likelihood $P(\bm{A}, \bm{T}, \bm{b}, \theta, \bm{c}, \gamma)=P(\bm{A},
\bm{T}| \bm{b}, \theta, \bm{c},
\gamma)P(\bm{b})P(\theta)P(\bm{c})P(\gamma)$ we can find the best
partition of the nodes and tags into groups, together with the number of
groups themselves, without overfitting. This happens because, in this
setting, the degrees of freedom of the model are themselves sampled from
a distribution, which will intrinsically ascribe higher probabilities to
simpler models, effectively working as a penalty on more complex
ones. An equivalent way of justifying this is to observe that the joint
likelihood can be expressed as $P(\bm{A}, \bm{T}, \bm{b}, \theta,
\bm{c}, \gamma)=2^{-\Sigma}$, where $\Sigma$ is the \emph{description
length} of the data, corresponding to the number of bits necessary to
encode both the data according to the model parameters as well as the
model parameters themselves. Hence, maximizing the joint Bayesian
likelihood is identical to the minimum description length (MDL)
criterion~\cite{grunwald_minimum_2007,
rosvall_information-theoretic_2007}, which is a formalization of Occam's
razor, where the simplest hypothesis is selected according to the
statistical evidence available.

We note that there are some caveats when selecting the priors
probabilities above. In the absence of \emph{a priori} knowledge, the
most straightforward approach is to select \emph{flat} priors that
encode this, and ascribe the same probability to all possible model
parameters~\cite{jaynes_probability_2003}. This choice, however, incurs
some limitations. In particular, it can be shown that with flat priors
it is not possible to infer with the SBM a number of groups that exceeds
an upper threshold that scales with $B_\text{max}\sim\sqrt{N}$, where
$N$ is the number of nodes in the
network~\cite{peixoto_parsimonious_2013}. Additionally, flat priors are
unlikely to be good models for real data, since they assume all
parameters values are equally likely. This is an extreme form of
randomness that encodes maximal ignorance about the model
parameters. However no data is truly sampled from such a maximally
random distribution; they are more likely to be sampled from some
nonrandom distribution, but with an unknown shape. An alternative,
therefore, is to postpone the decision on the prior until we observe the
data, by sampling the prior distribution itself from a hyperprior. Of
course, in doing so, we face the same problem again when selecting the
hyperprior. For the model at hand, we proceed in the following manner:
Since the matrices $\{e_{rs}\}$ and $\{m_{rs}\}$ are themselves
adjacency matrices of multigraphs (with $B_d$ and $B_d+B_t$ nodes,
respectively), we sample them from another set of SBMs, and so on,
following a nested hierarchy, until the trivial model with $B_d=B_t=1$
is reached, as described in Ref.~\cite{peixoto_hierarchical_2014}. For
the remaining model parameters we select only two-level Bayesian
hierarchies, since it can be shown that higher-level ones have only
negligible improvements asymptotically~\cite{peixoto_model_2015}. We
review and summarize the prior probabilities in
Appendix.~\ref{app:priors}. With this Bayesian hierarchical model, not
only we significantly increase the resolution limit to $B_\text{max}\sim
N/\ln N$~\cite{peixoto_hierarchical_2014}, but also we are able to
provide a description of the data at multiple scales.

It is important to emphasize that we are not restricting ourselves to
purely assortative structures, as it is the case in most community
detection literature, but rather we are open to a much wider range of
connectivity patterns that can be captured by the SBM. As mentioned in
the introduction, our approach differs from the parametric model
recently introduced by Newman and Clauset~\cite{newman_structure_2016},
where it is assumed that a node can connect to only one metadata tag,
and each tag is parametrized individually. In our model, a data node can
possess zero, one or more annotations, and the tags are clustered into
groups. Therefore our approach is suitable for a wider range of data
annotations, where entire classes of metadata tags can be
identified. Furthermore, since their approach is
parametric\footnote{More precisely, the approach of
Ref.~\cite{newman_structure_2016} is based on semi-Bayesian inference,
where priors for only part of the parameters are specified (the node
partition) but not others (the metadata-group and group-group
affinities, as well as node degrees). This approach is less susceptible
to overfitting when compared to pure maximum likelihood, but cannot be
used to select the model order (via the number of groups) as we do here,
for the reasons explained in the text (see also
Ref.~\cite{decelle_asymptotic_2011}).}, the appropriate number of groups
must be known beforehand, instead of being obtained from data, which is
seldom possible in practice. Additionally, when employing the fast MCMC
algorithm developed in Ref.~\cite{peixoto_efficient_2014}, the inference
procedure scales linearly as $O(N)$ (or log-linearly $O(N\ln^2N)$ when
obtaining the full hierarchy~\cite{peixoto_hierarchical_2014}), where
$N$ is the number of nodes in the network, independently of the number
of groups, in contrast to the expectation-maximization with belief
propagation of Ref.~\cite{newman_structure_2016}, that scales as
$O(B^2N)$, where $B$ is the number of groups being inferred. Hence, our
method scales well not only for large networks, but also for arbitrarily
large number of communities. An implementation of our method is freely
available as part of the graph-tool
library~\cite{peixoto_graph-tool_2014} at
\url{http://graph-tool.skewed.de}.

This joint approach of modelling the data and metadata allows us to
understand in detail the extent to which network structure and
annotations are correlated, in a manner that puts neither in advantage
with respect to the other. Importantly, we do not interpret the
individual tags as ``ground truth'' labels on the communities, and
instead infer their relationships with the data communities from the
entire data. Because the metadata tags themselves can be clustered into
groups, we are able to assess both their individual and collective
roles. For instance, if two tag nodes are assigned to the same group,
this means that they are both similarly informative on the network
structure, even if their target nodes are different. By following the
inferred probabilities between tag and node groups, one obtains a
detailed picture of their correspondence, that can deviate in principle
(and often does in practice) from the commonly assumed one-to-one
mapping~\cite{yang_defining_2012,hric_community_2014}, but includes it
as a special case.

\begin{figure*}
  \includegraphics[width=.8\textwidth]{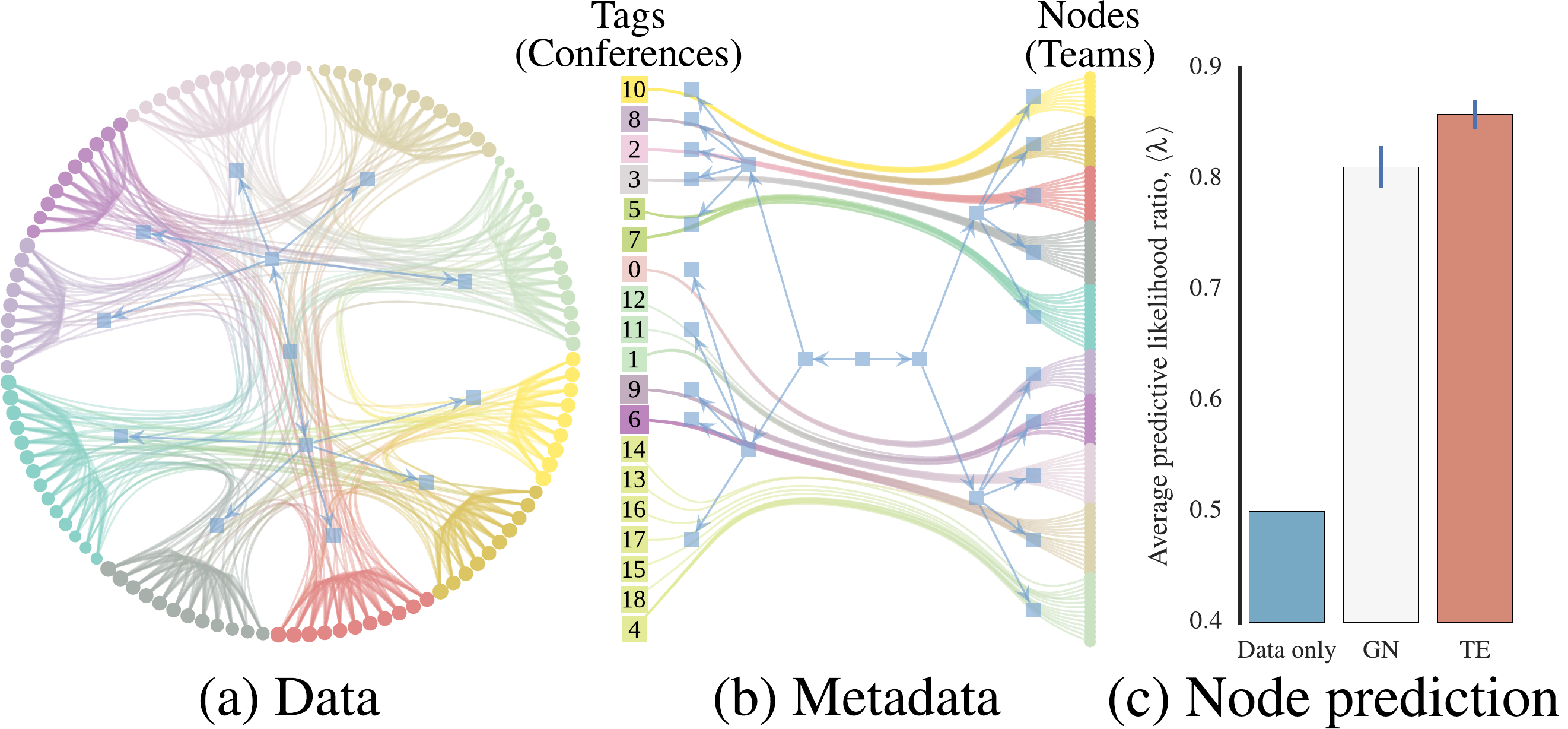}

  \caption{\label{fig:football}Joint data-metadata model inferred for
  the network of American football
  teams~\cite{girvan_community_2002}. (a) Hierarchical partition of the
  data nodes (teams), corresponding to the ``data'' layer. (b) Partition
  of the data (teams) and tag (conference) nodes, corresponding to the
  second layer. (c) Average predictive likelihood of missing nodes
  relative to using only the data (discarding the conferences), using
  the original conference assignment of
  Ref.~\cite{girvan_community_2002} (GN) and the corrected assignment of
  Ref.~\cite{evans_clique_2010} (TE).}
\end{figure*}

Before going into the systematic analysis of empirical datasets, we
illustrate the application of this approach with a simple example,
corresponding to the network of American college football
teams~\cite{girvan_community_2002}, where the edges indicate that a game
occurred between two teams in a given season. For this data it is also
available to which ``conferences'' the teams belong. Since it is
expected that teams in the same conference play each other more
frequently, this is assumed to be an indicator for the network
structure. If we fit the above model to this dataset, both the nodes
(teams) and tags (conferences) are divided into $B_d=10$ and $B_t=10$
groups, respectively (Fig.~\ref{fig:football}). Some of the conferences
correspond exactly to the inferred groups of teams, as one would
expect. However other conferences are clustered together, in particular
the independents, meaning that although they are collectively
informative on the network structure, individually they do not serve as
indicators of the network topology in a manner that can be conclusively
distinguished from random fluctuations.

In Fig.~\ref{fig:football} we used the conference assignments presented
in Ref.~\cite{evans_clique_2010}, which are different from the original
assignments in Ref.~\cite{girvan_community_2002}, due to a mistake in
the original publication, where the information from the wrong season
was used instead~\cite{evans_american_2012}.  We use this as an
opportunity to show how errors and noise in the metadata can be assessed
with our method, while at the same time we emphasize an important
application, namely the prediction of missing nodes. We describe it in
general terms, and then return to our illustration afterwards.

\subsection{Prediction of missing nodes}

To predict missing nodes, we must compute the likelihood of all edges
incident on it simultaneously, i.e. for an unobserved node $i$ they
correspond to the $i$th row of the augmented adjacency matrix,
$\bm{a}_i = \{A_{ij}'\}$, with $A_{kj}'=A_{kj}$ for $k \ne i$. If we
know the group membership $b_i$ of the unobserved node, in addition to
the observed nodes, the likelihood of the missing incident edges is
\begin{align}
  P(\bm{a}_i | \bm{A}, b_i, \bm{b}) &= \frac{\sum_\theta P(\bm{A}, \bm{a}_i | b_i, \bm{b}, \theta)P(\theta)}{\sum_\theta P(\bm{A} | \bm{b}, \theta) P(\theta)} \\
                           &= \frac{P(\bm{A}, \bm{a}_i | b_i, \bm{b}, \hat{\theta})P(\hat{\theta})}{P(\bm{A} | \bm{b}, \hat{\theta}') P(\hat{\theta}')},
\end{align}
where $\hat{\theta}$ and $\hat{\theta}'$ are the only choices of
parameters compatible with the node partition. However, we do not know
\emph{a priori} to which group the missing node belongs. If we have only
the network data available (not the metadata) the only choice we have is
to make the probability conditioned on the observed partition,
\begin{equation}\label{eq:pred_data}
  P(\bm{a}_i | \bm{A}, \bm{b}) = \sum_{b_i} P(\bm{a}_i | \bm{A}, b_i, \bm{b}) P(b_i|\bm{b}),
\end{equation}
where $P(b_i|\bm{b}) = P(\bm{b}, b_i)/P(\bm{b})$. This means that we can
use only the distribution of group sizes to guide the placement of the
missing node, and nothing more. However, in practical scenarios we may
have access to the metadata associated with the missing node. For
example, in a social network we might know the social and geographical
indicators (age, sex, country, etc) of a person for whom we would like
to predict unknown acquaintances. In our model, this translates to
knowing the corresponding edges in the tag-node graph $\bm{T}$. In this
case, we can compute the likelihood of the missing edges in the data
graph as
\begin{equation}
  P(\bm{a}_i|\bm{A}, \bm{T}, \bm{b},\bm{c}) = \sum_{b_i} P(\bm{a}_i|\bm{A},b_i,\bm{b}) P(b_i|\bm{T},\bm{b},\bm{c}),
\end{equation}
where the node membership distribution is weighted by the information
available in the full tag-node graph,
\begin{align}\label{eq:pred_metadata}
  P(b_i|\bm{T},\bm{b},\bm{c}) &= \frac{P(b_i, \bm{b}|\bm{T},\bm{c})}{P(\bm{b}|\bm{T},\bm{c})} \\
                              &= \frac{\sum_\gamma P(\bm{T}| b_i, \bm{b}, \bm{c}, \gamma) P(b_i, \bm{b})P(\gamma)}{\sum_{b_i'}\sum_\gamma P(\bm{T}|b_i', \bm{b}, \bm{c}, \gamma) P(b_i',\bm{b})P(\gamma)}\\
                              &= \frac{P(\bm{T}| b_i, \bm{b}, \bm{c}, \hat{\gamma}) P(b_i, \bm{b})P(\hat{\gamma})}{\sum_{b_i'}P(\bm{T}|b_i', \bm{b}, \bm{c}, \hat{\gamma}') P(b_i',\bm{b})P(\hat{\gamma}')},
\end{align}
where again $\hat{\gamma}$ and $\hat{\gamma}'$ are the only choices of
parameters compatible with the partitions $\bm{c}$ and $\bm{b}$. If the
metadata correlates well with the network structure, the above
distribution should place the missing node with a larger likelihood in
its correct group. In order to quantify the relative predictive
improvement of the metadata information for node $i$, we compute the
predictive likelihood ratio $\lambda_i\in[0,1]$,
\begin{equation}
  \lambda_i = \frac{P(\bm{a}_i|\bm{A}, \bm{T}, \bm{b},\bm{c})}{P(\bm{a}_i|\bm{A}, \bm{T}, \bm{b},\bm{c})+P(\bm{a}_i|\bm{A},\bm{b})},
\end{equation}
which should take on values $\lambda_i > 1/2$ if the metadata improves
the prediction task, or $\lambda_i < 1/2$ if it deteriorates it. The
latter can occur if the metadata \emph{misleads} the placement of the
node (we discuss below the circumstances where this can occur).

\begin{figure*}[ht!]
  \includegraphics[width=\textwidth]{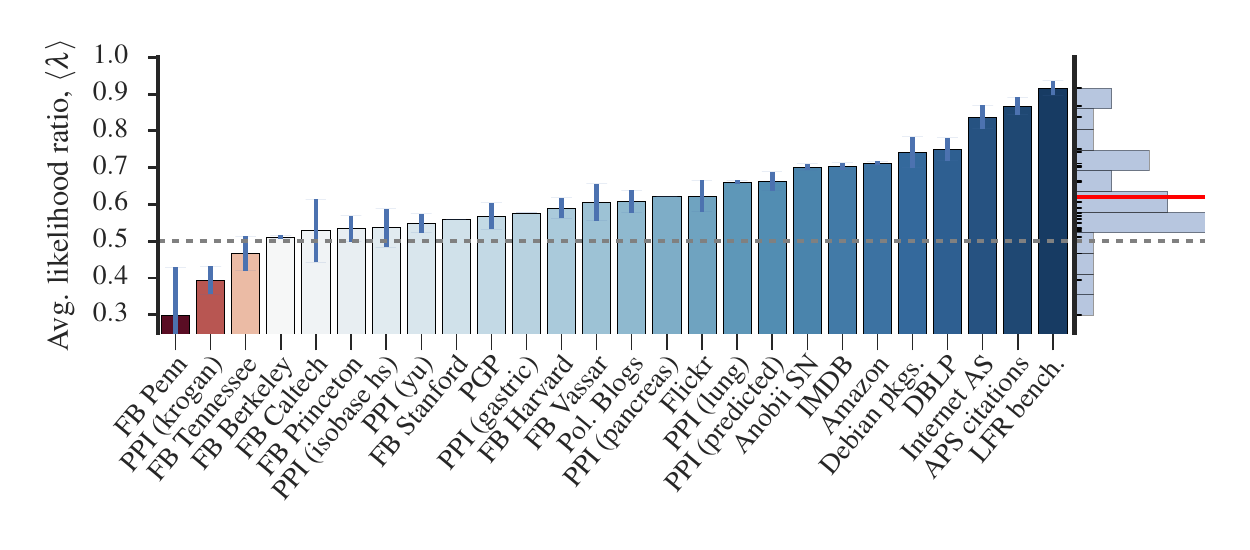}

  \caption{\label{fig:lambda-emp}
   Node prediction performance, measured by the average predictive
   likelihood ratio $\avg{\lambda}$ for a variety of annotated datasets
   (see Appendix~\ref{app:data} for descriptions).  Values above $1/2$
   indicate that the metadata improves the node prediction task. On the
   right axis a histogram of the likelihood ratios is shown, with a red
   line marking the average.}
\end{figure*}

In order to illustrate this approach we return to the American football
data, and compare the original and corrected conference assignments in
their capacity of predicting missing nodes. We do so by removing a node
from the network, inferring the model on the modified data, and
computing its likelihood according to Eq.~\ref{eq:pred_data} and
Eq.~\ref{eq:pred_metadata}, which we use to compute the average
predictive likelihood ratio for all nodes in the network,
$\avg{\lambda}=\sum_i\lambda_i/N$. As can be seen in
Fig.~\ref{fig:football}c, including the metadata improves the prediction
significantly, and indeed we observe that the corrected metadata
noticeably improves the prediction when compared to the original
inaccurate metadata. In short, knowing to which conference a football
team belongs, does indeed increase our chances of predicting against
which other teams it will play, and we may do so with a higher success
rate using the current conference assignments, rather than using those
of a previous year. These are hardly surprising facts in this
illustrative context, but the situation becomes quickly less intuitive
for datasets with hundreds of thousands of nodes and a comparable number
of metadata tags, for which only automated methods such as ours can be
relied upon.

\section{Empirical datasets}

We performed a survey of several network datasets with metadata
(described in detail in Appendix~\ref{app:data}), where we removed a
small random fraction of annotated nodes ($1\%$ or $100$ nodes,
whichever is smaller) many times, and computed the likelihood ratio
$\lambda_i$ above for every removed node. The average value for each
dataset is shown in Fig.~\ref{fig:lambda-emp}. We observe that for the
majority of datasets the metadata is capable of improving the prediction
of missing nodes, with the quality of the improvement being relatively
broadly distributed.  While this means that there is a positive and
statistically significant correlation between the metadata and the
network structure, for some datasets this leads only to moderate
predictive improvements. On the other hand, there is a minority of cases
where the inclusion of metadata \emph{worsens} the prediction task,
leading to $\avg{\lambda} < 1/2$. In such situations, the metadata seems
to divide the network in a manner that is largely orthogonal to the how
the network itself is connected. In order to illustrate this, we
consider some artificially generated datasets as follows, before
returning to the empirical datasets.

\subsection{Alignment between data and metadata}

We construct a network with $N$ nodes divided into $B_d$ equal-sized
groups, that are perfectly assortative, i.e. nodes of one group are only
connected to other nodes of the same group. Furthermore, the $E$ edges
of the network are randomly distributed among the groups, so that they
have on average the same edge density. This yields a simple structure
composed of the union of $B_d$ disjoint, fully random networks of similar density.

In the metadata layer we have the same number of $M=N$ metadata tags,
which are themselves also divided into an equal number $B_t=B_d=B$ of
equal-sized groups.

In order to place $E_m=E$ edges between data and metadata, we also
consider an \emph{alternative} partition $\{b'_i\}$ of the data nodes
into $B$ groups that is not equal to the original partition $\{b_i\}$
used to construct the network. A tag in one metadata group can only
connect randomly to nodes of one particular data group, and vice
versa. I.e. there is a one-to-one mapping between tag and data groups.

\begin{figure}
  \includegraphics{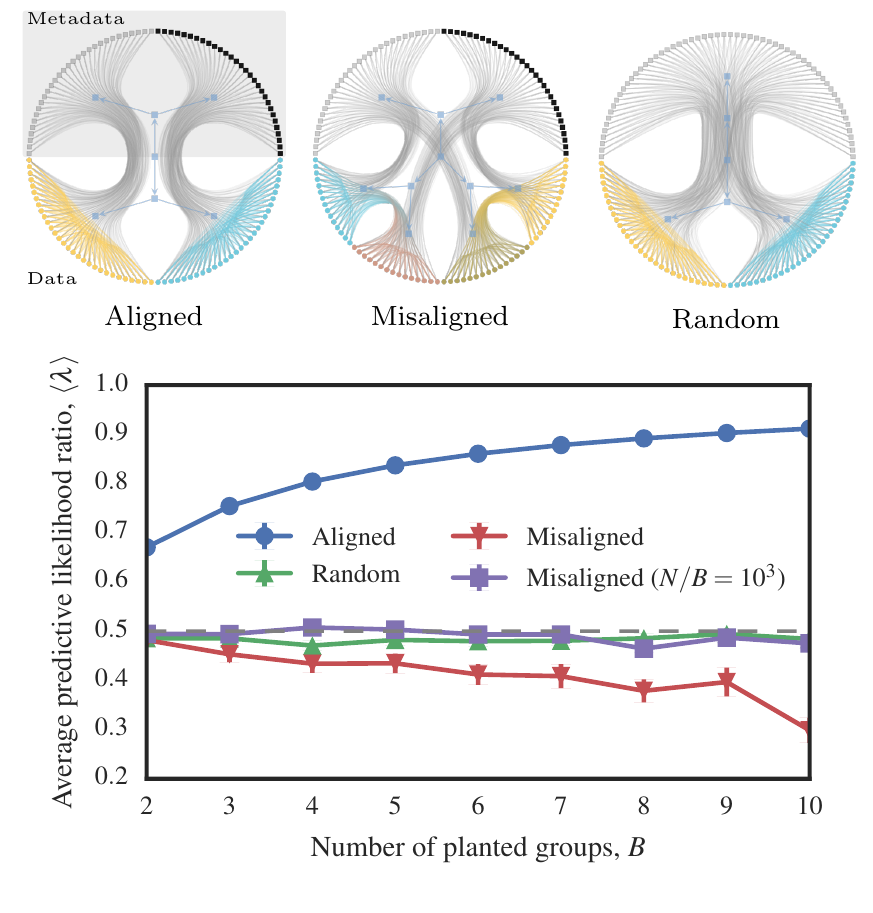}

  \caption{\label{fig:align}\emph{Top:} Examples of artificial annotated
  networks, showing aligned, misaligned and random metadata, as
  described in the text. \emph{Bottom:} Node prediction performance,
  measured by the likelihood ratio $\avg{\lambda}$, average over all
  possible single-node removals, for annotated networks generated with
  $B_d=B_t=B$ groups, $N=M=30\times B$ nodes and tags, $E=E_m=5\times N$
  node-node and tag-node edges, with specific network construction given
  by the legend. One of the curves corresponds to networks with
  misaligned metadata with a larger number of nodes, $N=M=10^3\times
  B$.}
\end{figure}

In total we consider three ways to connect the data with the
metadata:
\begin{enumerate}
\item \emph{Aligned} with the original data partition $\{b_i\}$,
      i.e. tag-node edges connect to the same data groups used to place
      the node-node edges;
\item  \emph{Misaligned} with the data partition, i.e. tag-node edges
      connect to the groups of the alternative data partition
      $\{b'_i\}$;
\item \emph{Random}: The tag-node edges are placed entirely at random,
      i.e.  respecting neither the tag nor the node partitions.
\end{enumerate}
We emphasize that 2 (misaligned) and 3 (random) are different: the
former corresponds to \emph{structured} metadata that is uncorrelated
with the network structure, and the latter corresponds to
\emph{unstructured} metadata. In other words, in the misaligned case the
node-tag graph is not fully random, since it only connects specific tag
groups to specific node groups, whereas in the random case the node-tag
edges are indeed fully random. An example of each type of construction
for $B=2$ is shown in Fig.~\ref{fig:align}.

When performing node prediction for artificial networks constructed in
this manner, one observes improved prediction with aligned metadata
systematically; however with misaligned metadata a measurable
degradation can be seen, while for random metadata neutral values close
to $\avg{\lambda}=1/2$ are observed (see Fig.~\ref{fig:align}). The
degradation observed for misaligned metadata is due to the subdivision
of the data groups into $B$ smaller subgroups, according to how they are
connected to the metadata tags. This subdivision, however, is not a
meaningful way of capturing the pattern of the node-node connections,
since all nodes that belong to the same planted group are statistically
indistinguishable. If the number of subgroups is sufficiently large,
this will invariably induce the incorporation of noise into the model
via the different number of edges incident on each
subgroup\footnote{Note that this incorporation of noise is not strictly
an overfitting, since the subdivisions are still required to properly
describe the data-metadata edges.}. Since these differences result only
from statistical fluctuations, they are bad predictors of unobserved
data, and hence cause the degradation in predictive quality. We note,
however, that in the limiting case where the number of nodes inside each
subdivision becomes sufficiently large, the degradation vanishes, since
these statistical fluctuations become increasingly less relevant (see
Fig.~\ref{fig:align}, curve $N/B = 10^3$). Nevertheless, for
sufficiently misaligned metadata the total number of inferred data
groups can increase significantly as $B_d=B_d^{0}\times B_t$, where
$B_d^{0}$ is the number of data groups used to generate the
network. Therefore, in practical scenarios, the presence of
\emph{structured} (i.e. non-random) metadata that is strongly
\emph{uncorrelated} with the network structure can indeed deteriorate
node prediction, as observed in a few of the empirical examples shown in
Fig.~\ref{fig:lambda-emp}.

\subsection{How informative are individual tags?}

The average likelihood ratio $\avg{\lambda}$ used above is measured by
removing nodes from the network, and include the simultaneous
contribution of all metadata tags that annotate them. However our model
also divides the metadata tags into classes, which allows us to identify
the predictiveness of each tag individually according to this
classification. With this, one can separate informative from
noninformative tags within a single dataset.

We again quantify the predictiveness of a metadata tag in its capacity
to predict which other nodes will connect to the one it
annotates. According to our model, the probability of some data node $i$
being annotated by tag $t$ is given by
\begin{equation}\label{eq:mpred}
  P_m^t(i | t) = d_i\frac{m_{b_i,c_t}}{m_{b_i}m_{c_t}},
\end{equation}
which is conditioned on the group memberships of both data and metadata
nodes. Analogously, the probability of some data node $i$
being a neighbor of a chosen data node $j$ is given by
\begin{equation}\label{eq:dpred}
  P_e(i | j) = k_i\frac{e_{b_i,b_j}}{e_{b_i}e_{b_j}}.
\end{equation}
Hence, the probability of $i$ being a neighbor of any node $j$ that is
annotated with tag $t$ is given by
\begin{equation}
  P_t(i) = \sum_j P(i | j) P_m(j | t).
\end{equation}
In order to compare the predictive quality of this distribution, we need
to compare it to a \emph{null} distribution where the tags connect
randomly to the nodes,
\begin{equation}
  Q(i) = \sum_j P(i | j) \Pi(j),
\end{equation}
where $\Pi(i) = d_i / M$, with $M=\sum_{r<s}m_{rs}$, is the probability
that node $i$ is annotated with any tag at random. The information gain
obtained with the annotation is then quantified by the Kullback-Leibler
divergence between both distributions,
\begin{equation}\label{eq:kl}
  D_{\text{KL}}(P_t||Q) = \sum_i P_t(i) \ln \frac{P_t(i)}{Q(i)}.
\end{equation}
This quantity measures the amount of information lost when we use the
random distribution $Q$ instead of the metadata-informed $P_t$ to
characterize possible neighbors, and hence the amount we gain when we
do the opposite. It is a strictly positive quantity, that can take any
value between zero and $-\ln Q^*$, where $Q^*$ is the smallest non-zero
value of $Q(i)$. If we substitute Eqs.~\ref{eq:dpred} and~\ref{eq:mpred}
in Eq.~\ref{eq:kl}, we notice that it only depends on the group
membership of $t$, and can be written as
\begin{equation}
  D_{\text{KL}}(P_t||Q) = D_{\text{KL}}(p_{c_t}||q)
\end{equation}
with
\begin{equation}
  p_r(u) = \sum_s p_e(u|s)p_m(s|u),\quad q(u) = \sum_s p_e(u|s)\pi(s),
\end{equation}
being the probabilities of a node that belongs to group $u$ being a
neighbor of a node annotated by a tag belonging to group $r$, for both
the structured and random cases, where $p_e(u|s) = e_{us}/e_{s}$,
$p_m(s|u) = m_{sr}/m_r$, and $\pi(s)=m_s / M$. Since this can take any
value between zero and $-\ln q^*$, where $q^*$ is the smallest non-zero
value of $q(u)$, this will in general depend on how many edges there are
in the network, given that $q^* \ge 1/2E$. For a concise comparison
between datasets of different sizes, it is useful to consider a relative
version of this measure that does not depend on the size. Although one
option is to normalize by the maximum possible value, here we use instead
the entropy of $q$, $H(q) = -\sum_rq(r)\ln q(r)$, and denote the
predictiveness $\mu_r$ of tag group $r$ as
\begin{equation}\label{eq:mu}
  \mu_r \equiv \frac{D_{\text{KL}}(p_r||q)}{H(q)}.
\end{equation}
This gives us the relative improvement of the annotated prediction with
respect to the uniformed one. Although it is possible to have $\mu_r >
1$, this is not typical even for highly informative tags, and would mean
that a particularly unlikely set of neighbors becomes particularly
likely once we consider the annotation. Instead, a more typical highly
informative metadata annotation simply narrows down the predicted
neighborhood to a \emph{typical} group sampled from $q$.

\begin{figure*}
  \includegraphics{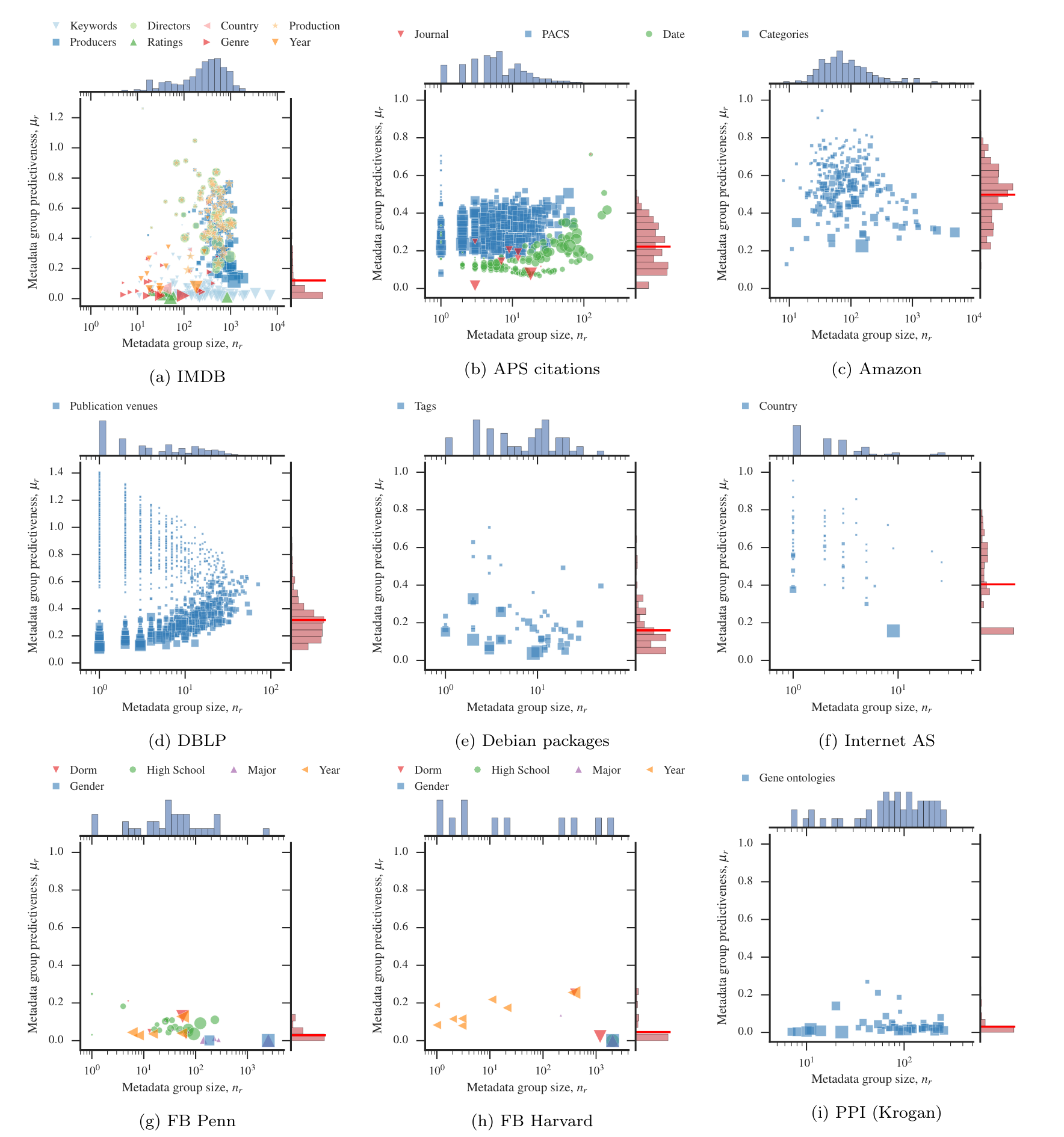}
  \caption{Metadata predictiveness for several empirical datasets.  The figures
    show the predictiveness of metadata groups $\mu_r$ (Eq.~\ref{eq:mu})
    vs. metadata group sizes, $n_r$.The sizes of the symbols indicate
    the metadata frequency. The symbols correspond to the most frequent
    types of tags in each group (which may contain tags of different
    types). On the axis of each figure are shown marginal histograms,
    weighted according to the tag frequencies. A red horizontal line
    marks the average predictiveness.\label{fig:pred}}
\end{figure*}

Using the above criterion we investigated in detail the datasets of
Fig.~\ref{fig:lambda-emp}, and quantified the predictiveness of the node
annotations, as is shown in Fig.~\ref{fig:pred} for a selected
subset. Overall, we observe that the datasets differ greatly not only in
the overall predictiveness of their annotations, but also in the
internal structures. Typically, we find that within a single dataset the
metadata predictiveness is widely distributed. A good example of this is
the IMDB data, which describes the connection between actors and films,
and includes annotations on the films corresponding to the year and
country of production, the producers, the production company, the
genres, user ratings as well as user-contributed keywords. In
Fig.~\ref{fig:pred}a we see that the larger fraction of annotations
posses very low predictiveness (which includes the vast majority of
user-contributed keywords and ratings), however there is still a
significant number of annotations that can be quite predictive. The most
predictive types of metadata are combinations of producers and directors
(e.g. Cartoon productions), followed by specific countries (e. g. New
Zealand, Norway) and year of productions. Besides keywords and ratings,
film genres are among those with the lowest predictiveness. A somewhat
narrower variability is observed for the APS citation data in
Fig.~\ref{fig:pred}b, where the three types of annotations are clearly
distinct. The PACS numbers are the most informative on average, followed
by the date of publication (with older dates being more predictive then
new ones --- presumably due to the increasing publication volume and
diversification over the years), and lastly the journal. One prominent
exception is the most predictive metadata group that corresponds to the
now-extinct ``Physical Review (Series I)'' journal, and its publication
dates ranging from 1893 to 1913. For the Amazon dataset of
Fig.~\ref{fig:pred}c, the metadata also exhibits significant predictive
variance, but there are no groups of tags that possess very low values,
indicating that most product categories are indeed strong indications of
co-purchases. This is similar to what is observed for the Internet AS,
with most countries being good predictors of the network structure. The
least predictive annotations happen to be a group of ten countries that
include the US as the most frequent one. A much wider variance is
observed in the DBLP collaboration network, where the publication venues
seem to be divided in two branches: very frequent and popular ones with
low to moderate predictiveness, and many very infrequent ones with high
to very high predictiveness. For other datasets a wide variance in
predictiveness is not observed. In particular for most Facebook networks
as well as protein-protein interaction networks, the available metadata
seems to be only tenuously correlated with the network structure, with
narrowly-distributed values of low predictiveness, in accordance with
their relatively low placement in Fig.~\ref{fig:lambda-emp}.

\section{Prediction of missing metadata}

Since we have defined a full joint model for data and metadata, our
framework is not restricted to prediction of missing nodes, but can also
predict missing edges both in the data and metadata layers. The latter
can be used to predict incomplete metadata information, which
corresponds to missing edges between data nodes and metadata tags, as
follows. Suppose the tag layer is decomposed as the union of two edge
sets, $\delta \bm{T} \cup \bm{T}$, where $\bm{T}$ is a set of observed
data-metadata edges, and $\delta \bm{T}$ is a set of missing edges of
the same type. Under our model, we can write the marginal posterior
likelihood for $\delta \bm{T}$ as
\begin{equation}\label{eq:marg_missing_tag}
  P(\delta\bm{T}|\bm{T}, b, c) = \frac{P(\delta\bm{T}\cup\bm{T}|\bm{T}, b, c)}{P(\bm{T}| b, c)},
\end{equation}
where $P(\bm{T}|b, c) = \sum_{\delta\bm{T}}P(\delta\bm{T}\cup\bm{T}|b,
c)$ is a normalization constant. If we have our set of missing edges
coming from a restricted set of possibilities, $\delta \bm{T} \in
\{\delta \bm{T}_1,\delta \bm{T}_2,\dots\}$, we may write the predictive
likelihood ratio
\begin{equation}\label{eq:missing_tag_lambda}
  \lambda_i = \frac{P(\delta\bm{T}_i|\bm{T}, b, c)}{\sum_jP(\delta\bm{T}_j|\bm{T}, b, c)} = \frac{P(\delta\bm{T}_i\cup\bm{T}| b, c)}{\sum_jP(\delta\bm{T}_j\cup\bm{T}| b, c)},
\end{equation}
where the normalization constant of Eq.~\ref{eq:marg_missing_tag} no
longer plays a role. Hence, if we want to compare the likelihood of a
given set of alternative node annotations, all we need to do is to infer
the parameters $b$ and $c$ of the model given the observed network,
\begin{equation}
  \{\hat{b}, \hat{c}\} = \underset{\{b,c\}}{\operatorname{arg max}}\; P(\bm{T}|b, c)P(b)P(c),
\end{equation}
and then add the missing edges $\delta\bm{T}_i$ to the likelihood using
this parameter estimate to compute the likelihood ratio of
Eq.~\ref{eq:missing_tag_lambda}.

\begin{figure}
  \includegraphics[width=\columnwidth]{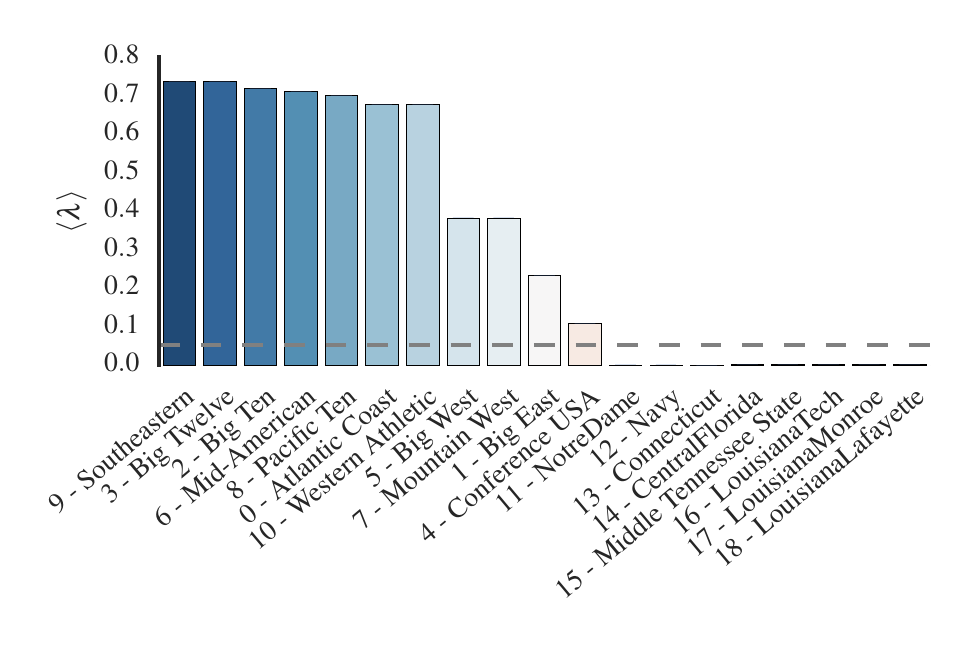} \caption{Average
  predictive likelihood ratio $\avg{\lambda}$ of missing metadata tags
  (conferences) for the American football data, using the annotations
  given in Ref.~\cite{evans_american_2012}. Tags $11$ to $18$ are
  ``independents'', i.e. teams that do not belong to any conference.
  The dashed line marks the value $1/19$, corresponding to a uniform
  likelihood between all tags.\label{fig:pred-tags}}
\end{figure}

We illustrate the application of our method again with the American
college football data. For each data node (team), we remove the single
metadata tag associated with it (i.e. the team's conference), perform
the model inference, and compute the predictive likelihood ratio of
Eq.~\ref{eq:missing_tag_lambda} for the removed tag, with respect to all
other possible tags. The averages over all teams that belong to a given
conference are shown in Fig.~\ref{fig:pred-tags}. The method succeeds in
predicting the correct conference assignment with the highest likelihood
in all cases, except for the ``independent'' teams. These teams do not
belong to any conference, and are therefore assigned a unique conference
tag. When this assignment is removed, it leaves an independent tag
without any connection to the graph, and hence out model is not able to
predict its placement. But since there is no additional information in
the data once this sole assignment is removed, it is simply impossible
to make an informative guess. In the cases where it is possible, our
approach seems able to leverage the available information and increases
the changes of successful metadata prediction.

\section{Conclusion}

We presented a general model for the large-scale structure of annotated
networks that does not intrinsically assume that there is a direct
correspondence between metadata tags and the division of network into
groups, or communities. Instead, we assume that the data-metadata
correlation is itself generated by an underlying process, with
parameters that are unknown \emph{a priori}. We presented a Bayesian
framework to infer the model parameters from data, which is capable of
uncovering --- in addition to the network structure --- the connection
between network structure and annotations, if there is one to be
found. We showed how this information can be used to predict missing
nodes in the network when only the annotations are known.

When applying the method for a variety of annotated datasets, we found
that their annotations lie in a broad range with respect to their
correlation with network structure. For most datasets considered, there
is evidence for statistically significant correlations between the
annotations and the network structure, in a manner that can be detected
by our method, and exploited for the task of node prediction. For a few
datasets, however, we found evidence of metadata which is not trivially
structured, but seems to be largely uncorrelated with the actual network
structure.

The predictiveness variance of metadata observed across different
datasets is also often found inside individual datasets. Typically,
single datasets possess a wealth of annotations, most of which are not
very informative on the network structure, but a smaller fraction
clearly is. Our method is capable of separating groups of annotations
with respect to their predictiveness, and hence can be used to prune
such datasets from ``metadata noise'', by excluding low-performing tags
from further analysis.

As is always true when doing statistical inference, results obtained are
conditioned on the validity of the model formulation, which invariably
includes assumptions about the data-generating process. In our case,
this means that the data-metadata layer can be represented as a graph,
and that it is well modelled by a SBM. Naturally, this is only one of
many possibilities, and it remains an open problem to determine which
alternatives work best for any given annotated network. This is
particularly true for annotations that correspond to continuous values
(e.g. time and space), which would need either to be discretized before
the application of our method, or preferably, would require a different
modelling ansatz (see
e.g. Ref.~\cite{newman_structure_2016}).

Nevertheless, we argue that the present approach is an appropriate
starting point, that provides an important but overlooked perspective in
the context of community detection validation. In a recent
study~\cite{hric_community_2014} a systematic comparison between various
community detection methods and node annotations was performed, where
for most of them strong discrepancies were observed. If we temporarily
(and unjustifiably) assume a direct agreement with available annotations
as the ``gold standard'', this discrepancy can be interpreted in a few
ways. Firstly, the methods might be designed to find structures that fit
the data poorly, and hence cannot capture their most essential
features. Secondly, even if the general ansatz is sound, a given
algorithm might still fail for more technical and subtle reasons. For
example, most methods considered in Ref.~\cite{hric_community_2014} do
not attempt to gauge the statistical significance of their results, and
hence are subject to
overfitting~\cite{guimera_modularity_2004,good_performance_2010}. This
incorporation of statistical noise will result in largely meaningless
division of the networks, which would be poorly correlated with the
``true'' division. Additionally, recently Newman and
Clauset~\cite{newman_structure_2016} suggested that while the
best-fitting division of the network can be poorly correlated with the
metadata, the network may still admit alternative divisions that are
also statistically significant, but happen to be well correlated with
the annotations.

On the other hand, the metadata heterogeneity we found with our method
gives a strong indication that node annotations should not be used in
direct comparisons to community detection methods in the first place
--- at least not indiscriminately. In most networks we analyzed, even
when the metadata is strongly predictive of the network structure, the
agreement between the annotations and the network division tends to be
complex, and very different from the one-to-one mapping that is more
commonly assumed. Furthermore, almost all datasets contain considerable
noise in their annotations, corresponding to metadata tags that are
essentially random. From this, we argue that data annotations should not
be used as a panacea in the validation of community detection
methods. Instead, one should focus on validation methods that are
grounded in statistical principles, and use the metadata as source of
additional evidence --- itself possessing its own internal structures
and also subject to noise, errors and omissions --- rather than a form
of absolute truth.

\begin{acknowledgments}
  We acknowledge the computational resources provided by the Aalto Science-IT project.
  D. H. and S.F. gratefully acknowledge MULTIPLEX, grant number 317532
  of the European Commission. T.P.P acknowledges support from the University of
  Bremen under funding program ZF04.
\end{acknowledgments}

\appendix

\section{Model likelihood and priors}\label{app:priors}

As mentioned in the text, the \emph{microcanonical} degree-corrected SBM
log-likelihood is given by~\cite{peixoto_entropy_2012}
\begin{equation}
  \ln P(\bm{A} | \bm{b}, \theta) \approx -E - \frac{1}{2}\sum_{rs}e_{rs}\ln\frac{e_{rs}}{e_re_s}-\sum_i\ln k_i!,
\end{equation}
(if Stirling's factorial approximation is used) and likewise for $\ln
P(\bm{T} | \bm{c}, \gamma)$, where one replaces $e_{rs}$ by $m_{rs}$ and
$k_i$ by $d_i$,
\begin{equation}
  \ln P(\bm{T} | \bm{c}, \gamma) \approx -M - \frac{1}{2}\sum_{rs}m_{rs}\ln\frac{m_{rs}}{m_rm_s}-\sum_i\ln d_i!,
\end{equation}
where $E = \sum_{rs}e_{rs}/2$ and $M = \sum_{rs}m_{rs}/2$. This assumes
that the graph is sufficiently sparse, otherwise corrections need to be
introduced, as described in Ref.~\cite{peixoto_entropy_2012,
peixoto_model_2015}. In order to compute the full joint likelihood, we
need priors for the parameters $\{b_i\}$, $\{c_i\}$, $\{k_i\}$,
$\{d_i\}$, $\{e_{rs}\}$ and $\{m_{rs}\}$.

For the node partitions, we use a two-level Bayesian hierarchy as done
in Ref.~\cite{peixoto_hierarchical_2014}, where one first samples the
group sizes from a random histogram, and then the node partition
randomly conditioned on the group sizes. The nonparametric likelihood is
given by $P(\{b_i\}) = e^{-\mathcal{L}_p}$, with
\begin{align}\label{eq:lp}
  \mathcal{L}_p = \ln{\textstyle\multiset{B}{N}} + \ln N! - \sum_r\ln n_r!,
\end{align}
where $\multiset{n}{m} = {n + m - 1 \choose m}$ is the total number of
$m$-combinations with repetitions from a set of size $n$. The prior
$P(\{c_i\})$ is analogous.

For the degree sequences, we proceed in the same
fashion~\cite{peixoto_model_2015}, by sampling the degrees conditioned
on the total number of edges incident on each group, by first sampling a
random degree histogram with a fixed average, and finally the degree
sequence conditioned on this distribution. This leads to a likelihood
$P(\{k_i\}|\{e_{rs}\},\{b_i\}) = e^{-\mathcal{L}_\kappa}$, with
\begin{equation}\label{eq:lkappa_D1}
  \mathcal{L}_{\kappa} = \sum_r\ln\Xi_r + \ln n_r! - \sum_k \ln n^r_k!,
\end{equation}
where $\ln\Xi_r \simeq 2\sqrt{\zeta(2)e_r}$.  Again, the likelihood for
$P(\{d_i\}|\{m_{rs}\},\{c_i\})$ is entirely analogous.

For the matrix of edge counts $\{e_{rs}\}$ we use the hierarchical prior
proposed in Ref.~\cite{peixoto_hierarchical_2014}. Here we view this
matrix as the adjacency matrix of a multigraph with $B_d$ nodes and
$E_d=\sum_{rs}e_{rs}/2$ edges. We sample this multigraph from another
SBM with a number of groups $B_d^{1}$, which itself is sampled from
another SBM with $B_d^{2}$ groups and so on, until $B_d^L=1$ for some
depth $L$.  The whole nonparametric likelihood is then $P(\{e_{rs}\}) =
e^{-\Sigma}$, with
\begin{equation}\label{eq}
  \Sigma = \sum_{l=1}^LS_m(\{e^l_{rs}\}, \{n^l_r\}) + \mathcal{L}^{l-1}_t,
\end{equation}
with $\{e^l_{rs}\}$, $\{n^l_r\}$ describing the block model at level
$l$, and
\begin{equation}\label{eq:sm}
  \mathcal{S}_m  = \sum_{r>s} \ln{\textstyle \multiset{n_rn_s}{e_{rs}}} + \sum_r \ln{\textstyle \multiset{\multiset{n_r}{2}}{e_{rr}/2}}
\end{equation}
is the entropy of the corresponding multigraph ensemble and
\begin{equation}\label{eq:dli}
  \mathcal{L}^l_t = \ln{\textstyle \multiset{B_l}{B_{l-1}}} + \ln B_{l-1}! - \sum_r \ln n_r^l!.
\end{equation}
is the description length of the node partition at level $l>0$. The
procedure is exactly the same for the prior $P(\{m_{rs}\})$.

\section{Datasets}\label{app:data}

Below we list descriptions of the annotated datasets used in this
work. Basic statistics are given in Table~\ref{tab:data}.

\begin{table}
\resizebox{\columnwidth}{!}{
\begin{tabular}{l|rr|rr|rr}
  \bf{Dataset}     &        $N_d$   &        $E_d$   &         $N_t$  &         $E_t$   &         $B_d$   &        $B_t$  \\
  \hline \hline
  LFR         &     1,000    &     9,839    &       40     &     1,000    &       29    &       29   \\
  PPI (Krogan)&     5,247    &    45,899    &     4,896    &    5,4904    &       62    &       55   \\
  PPI (Yu)    &      964     &     1,487    &     2,119    &    10,304    &       16    &       17   \\
  PPI (isobase-hs)& 8,580    &    34,250    &     1,972    &    20,633    &       40    &       15   \\
  PPI (gastric)&    4,763    &    26,131    &    10,445    &    94,035    &       50    &       50   \\
  PPI (lung)  &     4,843    &    27,459    &    10,948    &   100,492    &       55    &       50   \\
  PPI (pancreas)&   4,759    &    25,978    &    10,444    &    93,686    &       49    &       46   \\
  PPI (predicted)&  7,606    &    23,446    &    12,337    &   143,847    &       69    &       68   \\
  FB Caltech  &      762     &    16,651    &      591     &     4,145    &       22    &        5   \\
  FB Penn     &    41,536    & 1,362,220    &     4,805    &   216,349    &      365    &       29   \\
  FB Harvard  &    15,086    &   824,595    &     3,942    &    74,293    &      192    &       15   \\
  FB Stanford &    11,586    &   568,309    &     3,337    &    57,940    &      182    &       12   \\
  FB Berkeley &    22,900    &   852,419    &     2,906    &   116,556    &      267    &       16   \\
  FB Princeton&     6,575    &   293,307    &     2,396    &    32,901    &      110    &       10   \\
  FB Tennessee&    16,977    &   770,658    &     2,660    &    89,458    &      271    &       20   \\
  FB Vassar   &     3,068    &   119,161    &     1,620    &    16,859    &       69    &       12   \\
  Pol. blogs  &     1,222    &    16,714    &        2     &     1,222    &       12    &        2   \\
  DPD         &    35,029    &   161,313    &      580     &   115,999    &      253    &       59   \\
  PGP         &    39,796    &   197,150    &    35,370    &   148,966    &      485    &      380   \\
  Internet AS &    46,676    &   262,953    &      225     &    45,987    &      224    &       59   \\
  aNobii      &   140,687    &   869,448    &     8,003    &   926,403    &      194    &       70   \\
  Amazon      &   366,997    &   987,942    &    43,807    & 1,775,085    &    4,477    &      255   \\
  DBLP        &   317,080    &  1,049,866   &    13,477    &   719,820    &    4,667    &    1,746   \\
  IMDB        &   372,787    &  1,812,657   &   139,025    & 3,030,003    &      843    &      328   \\
  APS citations &  437,914   & 4,596,335    &    22,530    & 1,916,281    &    5,681    &      954   \\
  Flickr      &  1,624,992   & 15,476,836   &    99,270    & 8,493,666    &      779    &      158   \\
\end{tabular}
}
\caption{\label{tab:data}Summary of the basic statistics of the datasets
used in this work. $N_d$ and $E_d$ are the number of data nodes and
data-data edges, respectively, whereas $N_t$ and $E_t$ are the number of
metadata tags and node-tag edges, respectively. $B_d$ and $B_t$ are the
number of data and metadata groups inferred with our method.}
\end{table}

\paragraph{\bfseries LFR.}
Lancichinetti-Fortunato-Radicchi benchmark graph with $N=1000$ vertices
and community sizes between $10$ and $50$, with mixing parameter
$\mu=0.5$~\cite{lancichinetti_benchmark_2008}.  The remaining parameters
are the same as in Ref.~\cite{lancichinetti_benchmark_2008}. This model
corresponds to a specific parametrization of the degree-corrected
SBM~\cite{karrer_stochastic_2011}, and is often used to test and
optimize most current algorithms, and thus serves as a baseline
reference for a network with known and detectable structure.  The
network was created with standard LFR code available at
\url{https://sites.google.com/site/santofortunato/inthepress2}.

\paragraph{\bfseries PPI networks.}
In these networks nodes are individual proteins, and there is a link
between them if there is a confirmed interaction.  Protein labels from
Gene Ontology project (GO)\footnote{Retrieved from
\url{http://geneontology.org/.}} are used as node annotations.  The
networks themselves correspond to several different sources:
\emph{Krogan and Yu} correspond to yeast (Saccharomyces Cerevisiae),
from two different publications: Krogan~\cite{collins_toward_2007} and
Yu~\cite{yu_high-quality_2008}; \emph{isobase-hs} corresponds to human
proteins, as collected by the Isobase project~\cite{park_isobase:_2011};
\emph{Predicted} include predicted and experimentally determined
protein-protein interactions for humans, from the PrePPI
project~\cite{zhang_preppi:_2013} (human interactions that are in the HC
reference set predicted by structural modeling but not non-structural
clues); \emph{Gastric, pancreas, lung} are obtained by splitting the PrePPI
network~\cite{zhang_preppi:_2013} by the tissue where each protein is
expressed.

\paragraph{\bfseries Facebook networks (FB).}
Networks of social connections on the \url{facebook.com} online social
network, obtained in 2005, corresponding to students of different
universities~\cite{traud_social_2012}.  All friendships are present as
undirected links, as well as six types of annotation: Dorm (residence
hall), major, second major, graduation year, former high school, and
gender.

\paragraph{\bfseries Internet AS.}
Network of the Internet at the level of Autonomous Systems (AS).  Nodes
represent autonomous systems, i.e. systems of connected routers under
the control of one or more network operators with a common routing
policy.  Links represent observed paths of Internet Protocol traffic
directly from one AS to another. The node annotations are countries of
registration of each AS. The data were obtained from the CAIDA
project\footnote{http://www.caida.org/}.

\paragraph{\bfseries DBLP.}
Network of collaboration of computer scientists. Two scientists are
connected if they have coauthored at least one
paper~\cite{backstrom_group_2006}. Node annotations are publication
venues (scientific conferences).  Data is downloaded from
SNAP\footnote{Retrieved from
\url{http://snap.stanford.edu/data/com-DBLP.html}}~\cite{yang_defining_2012}.

\paragraph{\bfseries aNobii.}
This is an online social network for sharing book recommendations,
popular in Italy. Nodes are user profiles, and there can be two types of
directed relationships between them, which we used as undirected links
(``friends'' and ``neighbors'').  Data were provided by Luca
Aiello~\cite{aiello_people_2012,aiello_link_2010}.  We used all present
node metadata, of which there are four kinds: Age, location, country,
and membership.

\paragraph{\bfseries PGP.}
The ``Web of trust'' of PGP (Pretty Good Privacy) key signings,
representing an indication of trust of the identity of one person
(signee) by another (signer).  A node represents one key, usually but
not always corresponding to a real person or organization.  Links are
signatures, which by convention are intended to only be made if the two
parties are physically present, have verified each others' identities,
and have verified the key fingerprints. Data is taken from a 2009
snapshot of public SKS keyservers~\cite{richters_trust_2011}.

\paragraph{\bfseries Flickr.}
Picture sharing web site and social network, as crawled by Mislove et
al~\cite{mislove_measurement_2007}.  Nodes are users and edges exist if
one user ``follows'' another. The node annotations are  user groups centered
around a certain type of content, such as ``nature'' or
``Finland''.

\paragraph{\bfseries Political Blogs.}
A directed network of hyperlinks between weblogs on US politics,
recorded in 2005 by Adamic and Glance~\cite{adamic_political_2005}.
Links are all front-page hyperlinks at the time of the crawl. Node
annotations are ``liberal'' or ``conservative'' as assigned by either
blog directories or occasional self-evaluation.

\paragraph{\bfseries Debian packages.}
Software dependencies within the Debian GNU/Linux operating
system\footnote{\url{http://www.debian.org}}.  Nodes are unique
software packages, such as \texttt{linux-image-2.6-amd64},
\texttt{libreoffice-gtk}, or \texttt{python-scipy}.  Links are the
``depends'', ``recommends'', and ``suggests'' relationships, which are a
feature of Debian's APT package management system designed for tracking
dependencies.  Node annotations are tag memberships from the DebTags
project\footnote{\url{https://wiki.debian.org/Debtags}}, such as
\texttt{devel::lang:python} or \texttt{web::browser}~\cite{zini_cute_2005}.  The
network was generated from package files in Debian 7.1 Wheezy as of
2013-07-15, ``main'' area only.  Similar files are freely available in
every Debian-based OS.  Tags can be found in the \texttt{*\_Packages}
files in the \texttt{/var/lib/apt/} directory in an installed system or
on mirrors, for example
\url{ftp://ftp.debian.org/debian/dists/wheezy/main/binary-amd64/}.

\paragraph{\bfseries amazon.}
Network of product copurchases on online retailer \url{amazon.com}.
Nodes represent products, and edges are said to represent copurchases by
other customers presented on the product
page~\cite{leskovec_dynamics_2007}.  The true meaning of links is
unknown and is some function of Amazon's recommendation algorithm.  Data
was scraped in mid-2006 and downloaded from
\url{http://snap.stanford.edu/data/amazon-meta.html}.  We used
copurchasing relationships as undirected edges.
Product categories were used as node annotations. Although product
categories are hierarchical by nature, we used only the endpoints (or
``leaves'') of the hierarchy: \texttt{Books/Fiction/Fantasy/Epic} and
\texttt{Books/Nonfiction} are two different metadata labels.

\paragraph{\bfseries IMDB.}
This network is compiled by extracting information available in the
Internet Movie Database (IMDB)\footnote{\url{http://www.imdb.com}}, and
it contains each cast member and film as distinct nodes, and an
undirected edge exists between a film and each of its cast members. The
network used here corresponds to a snapshot made in
2012~\cite{peixoto_parsimonious_2013}. The node annotations are the
following information available on the films: Country and year of
production, production company, producers, directors, genre,
user-contributed keywords and genres.

\paragraph{\bfseries APS citations.}
This network corresponds to directed citations between papers published
in journals of the American Physical Society for a period of over 100
years\footnote{Retrieved from \url{http://publish.aps.org/dataset}}. The
node annotations correspond to PACS classification tags, journal and
publication date.

\bibliography{bib}

\end{document}